\documentclass[journal,10pt,doublecolumn]{IEEEtran}
\usepackage{mathtools, cuted}
\usepackage{lipsum, color}

\usepackage{footnote}
\usepackage[table]{xcolor}
\usepackage{times}
\usepackage{epsfig}
\usepackage{amsmath}
\usepackage{amsthm}
\usepackage{amsfonts}
\usepackage{graphicx}
\usepackage{amssymb}
\usepackage{amstext}
\usepackage{latexsym}
\usepackage{color,colortbl}
\usepackage{ifthen}
\usepackage{multirow}
\usepackage{verbatim}
\usepackage{array,tabularx}
\usepackage{arydshln}
\usepackage[mathscr]{euscript}
\usepackage{accents}
 \usepackage{cite}
\usepackage{hhline}
\usepackage{caption}
\usepackage{subcaption}
\usepackage{enumerate}
\usepackage{xcolor}
\usepackage{mathtools}
\usepackage{url}
\usepackage{xparse}
\usepackage{makecell}
\usepackage{varwidth}
\usepackage{bm}
\usepackage{arydshln}
\usepackage{footmisc} % footref

\usepackage{comment}
\usepackage{wrapfig}

\usepackage{caption}
\usepackage{float}
\usepackage{booktabs}

\usepackage{mathtools}

\usepackage{booktabs}

\usepackage{float}

\usepackage{multirow}

\usepackage{algorithm,algpseudocode}

\newcolumntype{C}[1]{>{\centering\let\newline\\\arraybackslash\hspace{0pt}}m{#1}}

\usepackage[normalem]{ulem}

\usepackage{amsmath,pgfplots,amssymb}
\usepackage{subcaption,graphicx}

\theoremstyle{definition}

\theoremstyle{definition}

\theoremstyle{definition}

\newcommand{\off}[1]{}

\usetikzlibrary{matrix,decorations.pathreplacing}
\usetikzlibrary{arrows.meta}

\definecolor{DarkGreen}{rgb}{0.1,0.5,0.1}
\definecolor{DarkRed}{rgb}{0.5,0.1,0.1}
\definecolor{DarkBlue}{rgb}{0.1,0.1,0.5}
\definecolor{DarkPurple}{rgb}{0.5,0.2,0.5}
\definecolor{DarkTurquoise}{rgb}{0.1,0.5,0.5}

\definecolor{beaublue}{rgb}{0.74, 0.83, 0.9}
\definecolor{coolblack}{rgb}{0.0, 0.18, 0.39}
\definecolor{apricot}{rgb}{0.98, 0.81, 0.69}
\definecolor{burntorange}{rgb}{0.8, 0.33, 0.0}
\definecolor{blue-violet}{rgb}{0.54, 0.17, 0.89}
\definecolor{byzantium}{rgb}{0.44, 0.16, 0.39}
\definecolor{brilliantrose}{rgb}{1.0, 0.33, 0.64}
\definecolor{cerisepink}{rgb}{0.93, 0.23, 0.51}
\definecolor{cobalt}{rgb}{0.0, 0.28, 0.67}
\definecolor{bostonuniversityred}{rgb}{0.8, 0.0, 0.0}

\algnewcommand\algorithmicforeach{\textbf{for each}}
\algdef{S}[FOR]{ForEach}[1]{\algorithmicforeach\ #1\ \algorithmicdo}

\ifCLASSINFOpdf
\else
\fi
\begin{document}
%\title{Lifting Private Information Retrieval}
\title{Post-Quantum Security for Ultra-Reliable Low-Latency Heterogeneous Networks}

\author{%
   \IEEEauthorblockN{Rafael G. L. D’Oliveira\IEEEauthorrefmark{1},
                     Alejandro Cohen\IEEEauthorrefmark{1},
                     John Robinson\IEEEauthorrefmark{2},
                    Thomas Stahlbuhk\IEEEauthorrefmark{3},
                     and Muriel M\'edard\IEEEauthorrefmark{1}}\\
   \IEEEauthorblockA{\IEEEauthorrefmark{1}%
                     RLE, MIT, Cambridge, MA, USA, \{rafaeld, cohenale, medard\}@mit.edu}\\
                     \IEEEauthorblockA{\IEEEauthorrefmark{2}%
                     University of North Georgia, GA, USA, johnny10robinson@gmail.com }\\
   \IEEEauthorblockA{\IEEEauthorrefmark{3}%
                     Lincoln Laboratory, MIT, Lexington, MA, USA,  thomas.stahlbuhk@ll.mit.edu}\\
\vspace{-0.8cm}}

\maketitle


\maketitle

%%%%%%%%%%%%%%%%%%%%%%%%%%%%%%%%%%%%%%%%%%%%
\begin{abstract}
%%%%%%%%%%%%%%%%%%%%%%%%%%%%%%%%%%%%%%%%%%%%
We consider the problem of post-quantum secure and ultra-reliable communication through a heterogeneous network consisting of multiple connections. Three performance metrics are considered: security, throughput, and in-order delivery delay. In this setting, previous work has looked, individually, at the trade-offs between in-order delivery delay and throughput, and between security and throughput. This is the first work considering the trade-off between all three for heterogeneous communication networks, while taking the computational complexity into account. We present LL-HUNCC, a low latency hybrid universal network coding cryptosystem. LL-HUNCC is an efficient coding scheme which allows for secure communications over a noisy untrusted heterogeneous network by encrypting only a small part of the information being sent. This scheme provides post-quantum security with high throughput and low in-order delivery delay guarantees. We evaluate LL-HUNCC via simulations on a setting inspired by a practical scenario for heterogeneous communications involving a satellite communication link and a 5G communication network. Under this scenario, we compare LL-HUNCC to the state-of-the-art where all communication paths are encrypted via a post-quantum public-key cryptosystem.

%%%%%%%%%%%%%%%%%%%%%%%%%%%%%%%%%%%%%%%%%%%%
\end{abstract}
%%%%%%%%%%%%%%%%%%%%%%%%%%%%%%%%%%%%%%%%%%%%

%%%%%%%%%%%%%%%%%%%%%%%%%%%%%%%%%%%%%%%%%%%%
  \section{Introduction}
%%%%%%%%%%%%%%%%%%%%%%%%%%%%%%%%%%%%%%%%%%%%

\begin{figure}[t]
    \centering
    \includegraphics[width = 1\columnwidth]{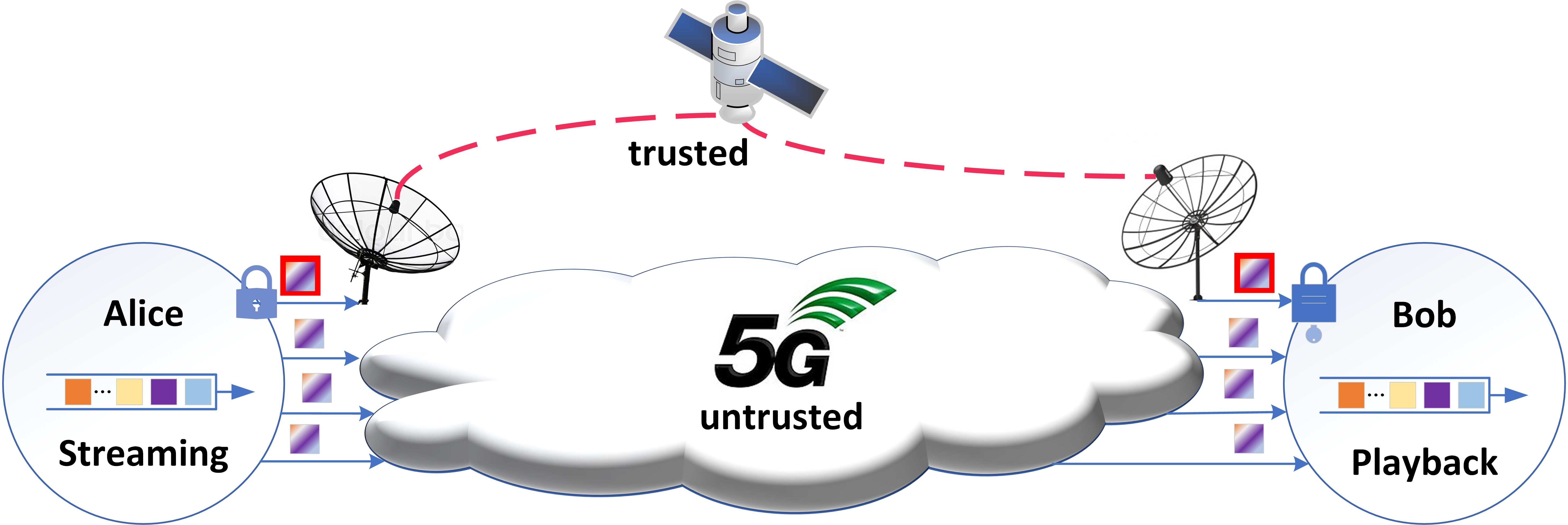}
    \caption{A heterogeneous communication network between Alice (the transmitter) and Bob (the receiver). Encryption by a public-key cryptosystem may only be performed on a subset of the communication links, which we refer to as trusted links. Our goal is to obtain post-quantum secure and reliable communication between Alice and Bob, while minimizing the in-order delivery delay and maximizing the throughput.}
    \label{fig:use_case}
    \vspace{-0.3cm}
\end{figure}

We consider the problem of post-quantum secure and ultra-reliable communication through a heterogeneous network consisting of multiple connections. In this problem, Alice (the transmitter) wishes to send a private message to Bob (the intended receiver) in the presence of eavesdropper, Eve. The communication network, illustrated in Fig. \ref{fig:use_case}, consists of $\ell$ noisy communication links, each modeled as an independent binary erasure channel (BEC). We assume that encryption, by a public-key cryptosystem, may only be performed on a subset of the communication links, which we refer to as trusted links. As is usual in these networks, we make use of a feedback channel. Our goal is to securely transmit confidential data over the noisy communication links, in the presence of a strong eavesdropper which may observe all the information in the network, while minimizing the in-order delivery delay and maximizing the throughput.

The non-secure version of this problem has been effectively handled, in terms of maximizing the throughput, via block Random Linear Network Coding (RLNC) \cite{koetter2003algebraic}. The main idea in this approach is to transmit random linear combinations of the messages over a large enough field. Then, once Bob is able to obtain a sufficient amount of these linear combinations, he can decode the message by solving a linear system. This approach achieves the channel capacity for multipath and multi-hop (MP-MH) networks. However, this occurs in the large block-length regime, and thus, does not guarantee a low in-order delivery delay which is critical in many applications.

In order to obtain the desired trade-off between in-order delivery delay and throughput, an Adaptive and Causal RLNC (AC-RLNC) joint scheduling-coding algorithm was recently presented, for both a single-path (SP) link \cite{9076631} and an MP-MH network \cite{9245536}. The AC-RLNC solution is adaptive to the channel condition and is causal in the sense that the data transmissions from the source depend on the particular realizations of the channel state, as reflected in the feedback acknowledgments from the destination. Thus, the algorithm adjusts the forward error correction re-transmission of RLNC packets to achieve the desired throughput vs. delay trade-off, as required for the particular application. The proposed re-transmission criterion tracks the actual network packet receiving rate and the missing Degree of Freedom (DoF) rate required at the destination to decode the received coded packets.

For the secure version of the problem, many secure network codes have been proposed \cite{bloch2011physical}. Although these schemes offer information-theoretic security, and are therefore post-quantum secure, they suffer from two issues. First, they assume that the eavesdropper is weak, in the sense that she does not observe all the information transmitted across the network. And second, their security comes with significant costs to the throughput.

% For the secure version on the problem, many secure RLNC codes have been proposed in the literature \cite{cai2011secure,el2012secure}. However, for the case of a strong eavesdropper, i.e., one which can observer all the information in the network, these secure codes do not guarantee post-quantum security of the information transmitted across the network with high data rates. In the case of a weak eavesdropper, i.e., one which observes a partial amount of the information in the network, traditional secure codes offer security at the price of throughput \cite{wyner1975wire,bloch2011physical}.

\begin{figure*}[t]
    \centering
    \includegraphics[scale= 0.4]{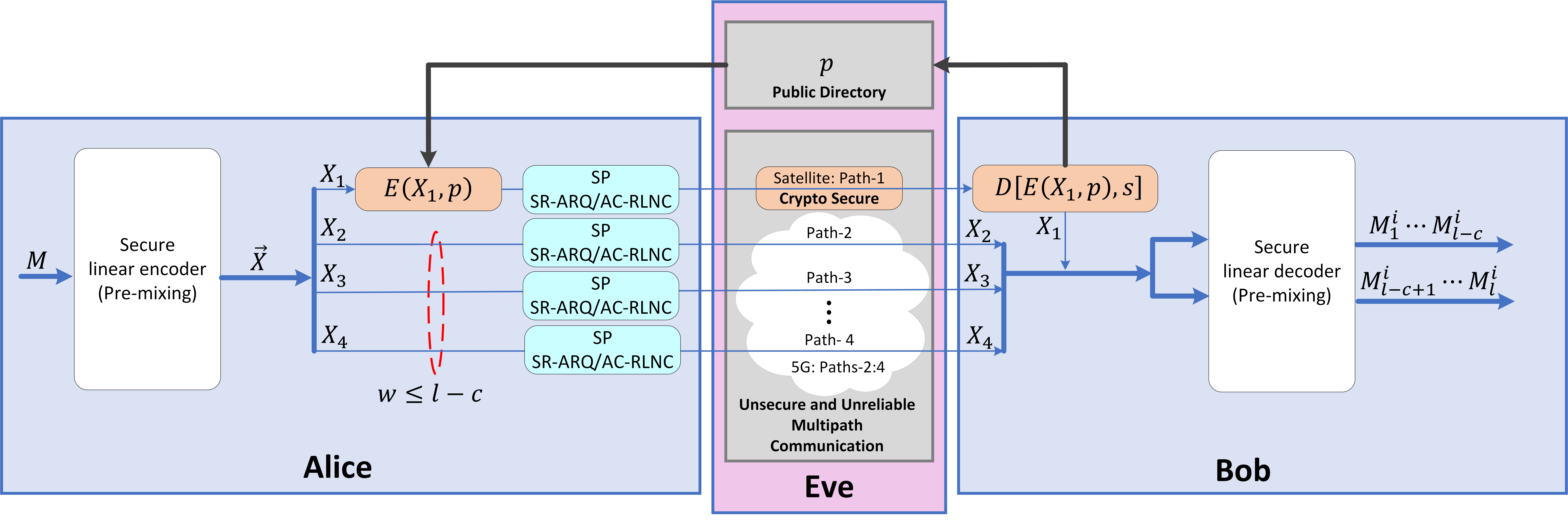}
    \caption{The LL-HUNCC scheme for $\ell = 4$ communication links, $c=1$ of which we apply encryption. The scheme consists of a security stage and a reliability stage. In the security stage Alice's message $M$ is first encoded by a secure code which outputs individually secure encodings $X_1, \ldots, X_4$. She then encrypts the first encoding into $E(X_1,p)$ utilizing a post-quantum McEliece cryptosystem. In the reliability stage, Alice chooses one of three approaches: SP SR-ARQ , SP AC-RLNC or MP AC-RLNC. Each approach provides different trade-offs between security, throughput, and in-order delivery delay. After receiving enough packets, Bob can retrieve the encoded messages, decrypt $X_1 = D(E(X_1,p),s)$, and finally, decode the original messages.}
    \label{fig:algo_model}
\end{figure*}

A secure scheme, which obtains a trade-off between a post-quantum security guarantee (for a strong Eve) and throughput, was presented in \cite{cohen2021network}. The scheme is called a hybrid universal network-coding cryptosystem (HUNCC). It is hybrid in that it combines information-theoretic security with a public-key cryptosystem, and is universal in that it can be applied to any communication system. HUNCC works by first premixing the data using a particular type of secure network coding scheme \cite{cohen2018secure} and then encrypting a small part of the mixed data before transmitting across the (noiseless) untrusted network.\footnote{Thus, HUNCC offers information-theoretic security against a weak Eve and post-quantum computational security guarantees against a strong Eve.} This allows for the throughput to approach one as the amount of data used for the premixing stage grows. However, this increase in the data used for premixing comes at a cost of a higher in-order delivery delay. Unlike the non-secure AC-RLNC solution where a sufficient amount of DoF at the destination allows Bob to decode the information in-order and immediately, with HUNCC, Bob needs all the packets from the premixing stage together in order to decode.

Thus, as elaborated above, previous work has looked individually at the trade-offs between in-order delivery delay and throughput, and between security and throughput. In this paper we consider the trade-off between all three of these quantities. We also present low latency HUNCC (LL-HUNCC), an efficient coding scheme with two independent stages, illustrated in Fig. \ref{fig:algo_model}. The first stage is a pre-processing stage in which we code for security and in the second stage we code for reliability. For security, we consider the linear mixing performed by HUNCC \cite{cohen2021network}, where a post-quantum McEliece encryption \cite{mceliece1978public} is applied to the trusted communication links. For reliability, we consider three options: SP SR-ARQ \cite{weldon1982improved}, SP AC-RLNC \cite{9076631} and MP AC-RLNC \cite{9245536}. These choices provide different trade-offs between security, throughput, and in-order delivery delay.

To evaluate these trade-offs we focus on a practical scenario for heterogeneous communications consisting of a trusted satellite communication link (e.g., LEO, MEO, or GEO satellite systems) and an untrusted 5G network. We note that, although the 5G communication is untrusted, it can significantly increase the efficiency of transmitting data information in terms of maximizing throughput and minimizing in-order delivery delays apart from reducing the cost of using the scarce bandwidth on the satellite link as opposed to the more ready access to bandwidth in 5G networks. We perform simulations for this scenario in Section \ref{sec: Evaluation and Discussion} and compare LL-HUNCC to the state-of-the-art in which all the communication links are encrypted by a post-quantum McEliece cryptosystem. 

In our experimental evaluations we observe that LL-HUNCC outperforms the state-of-the-art in terms of throughput for all second stage choices. However, in terms of in-order delivery delay of packets (relevant to real-time control applications), LL-HUNCC may suffer a degradation when compared to the state-of-the-art if it uses SP SR-ARQ or SP AC-RLNC in its second stage, but has similar performance if MP AC-RLNC is used instead. For the cases of in-order delivery delay of frames (relevant to video streaming applications), and in-order delivery delay of files (relevant to distributed storage applications), the differences between SP SR-ARQ, SP AC-RLNC, and MP AC-RLNC are negligible. Thus, overall we observe that LL-HUNCC is able to significantly outperform the state-of-the-art in terms of throughput at a negligible cost of the in-order delivery delay.

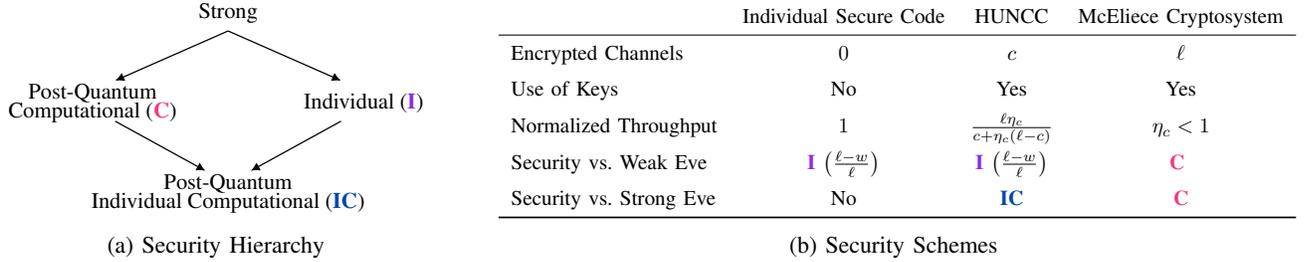
\begin{figure*}[!t]
\begin{subfigure}[]{0.39\textwidth}
    \centering
    \resizebox{0.8\textwidth}{!}{\begin{tikzpicture}[scale=1]

% \draw [rounded corners] (0,0) rectangle (10,5);

\node[align=center] at (5,5) {\Large{Strong}};

\draw[-{Latex[length=2mm, width=2mm]}, color=black, thick] (5,4.6) -- (7.5,3.5);

\draw[-{Latex[length=2mm, width=2mm]}, color=black, thick] (5,4.6) -- (2.5,3.5);

\node[align=center] at (8,3) {\Large{Individual} ({\color{blue-violet}\textbf{I}})};

\node[align=center] at (2,3) {\Large{Post-Quantum} \\ \Large{Computational} ({\color{cerisepink}\textbf{C}})};

\draw[-{Latex[length=2mm, width=2mm]}, color=black, thick] (2.5,2.6) -- (4.5,1.5);

\draw[-{Latex[length=2mm, width=2mm]}, color=black, thick] (7.5,2.6) -- (5.5,1.5);

\node[align=center] at (5,1) {\Large{Post-Quantum} \\ \Large{Individual Computational} ({\color{cobalt}\textbf{IC}})};

\end{tikzpicture}}
    \caption{Security Hierarchy}
    \label{fig: security and schemes a}
\end{subfigure}
\begin{subfigure}[]{0.59\textwidth}
    \centering
    \resizebox{\textwidth}{!}{
    \begin{tabular}{l c c c c c c}
    \toprule
         & Individual Secure Code  & HUNCC & McEliece Cryptosystem \\ \midrule
         Encrypted Channels & $0$  & $c$ & $\ell $\\[5pt]
         
        Use of Keys  & No & Yes & Yes\\[5pt]
         Normalized Throughput & $1$ & $\frac{\ell  \eta_c}{c+ \eta_c (\ell -c)}$ & $\eta_c < 1$\\[5pt]
         
         Security vs. Weak Eve  & {\color{blue-violet}\textbf{I}} $\left(\frac{\ell -w}{\ell }\right)$ & {\color{blue-violet}\textbf{I}} $\left(\frac{\ell -w}{\ell }\right)$ & {\color{cerisepink}\textbf{C}} \\[5pt]
         
         Security vs. Strong Eve & No  & {\color{cobalt}\textbf{IC}} & {\color{cerisepink}\textbf{C}} \\
        \bottomrule
\end{tabular}}
    \caption{Security Schemes}
    \label{fig: security and schemes b}
\end{subfigure}
\caption[lof]{\textbf{(a)} A schematic illustrating the hierarchy of security notions used in this paper. Arrows denote security implications, i.e. while strong security implies in both post-quantum computational and individual security, the latter two are not directly comparable. We note that individual security is specified in terms of a security parameter $\textbf{I}(\alpha)$ for $0\leq \alpha \leq 1$, with $\textbf{I}(0)$ corresponding to no security and $\textbf{I}(1)$ to strong security.  \textbf{(b)} Characteristics of the three security schemes relevant to this paper. We note that HUNNC obtains a better throughput than a pure MCEliece cryptosystem by loosening the security notion in the case of a strong eve. However, in the case of a weak Eve, HUNCC is able to guarantee information-theoretic security.}
\label{fig: security and schemes}
\end{figure*}
 
%%%%%%%%%%%%%%%%%%%%%%%%%%%%%%%%%%%%%%%%%%%%
% \subsection{Related Works}

% For single path (1 link) delay trouput trade off (Adapative causal network coding with feedback)

% not too many for multipath (multihop )

% Security take from huncc

%%%%%%%%%%%%%%%%%%%%%%%%%%%%%%%%%%%%%%%%%%%%
\subsection{Main Contributions}

To the best of our knowledge, this is the first work evaluating the trade-off between post-quantum security, throughput, and in-order delivery delay, for heterogeneous communication networks, while considering computational complexity. We present LL-HUNCC, an efficient coding scheme which allows for low-latency secure communications over a noisy untrusted heterogeneous network by encrypting only a small part of the information being sent. We evaluate LL-HUNCC in a practical setting which simulates a heterogeneous network consisting of a trusted satellite communication and an untrusted 5G network. In our simulations we observe that LL-HUNCC significantly outperforms the state-of-the-art in terms of throughput at a negligible cost to the in-order delivery delay.

%%%%%%%%%%%%%%%%%%%%%%%%%%%%%%%%%%%%%%%%%%%%
\section{Security, Throughput, and Delay}
%%%%%%%%%%%%%%%%%%%%%%%%%%%%%%%%%%%%%%%%%%%%

In this section, we provide the definitions of our main performance metrics: security, throughput, and delay.

The \emph{throughput} of a communication scheme is the total amount of data information, in units of bits per second, which are delivered to Bob. We generally focus on the \emph{normalized throughput}, denoted by $\eta$, which is the total amount of data information delivered to Bob divided by the total amount of bits sent to him. E.g., if to communicate a $10$-bit message $M$, Alice must send two $15$-bits packets $X_1$ and $X_2$, then the normalized throughput of this scheme is $\eta = \frac{10}{15+15}=\frac{1}{3}$.

Our next notion is that of the in-order delivery delay. We consider three versions based on the following hierarchy: An information packet is a sequence of bits which are sent together through the communication network. A frame of size $k$ is a collection of $k$ information packets. And finally, a file is a collection of frames.

The \emph{in-order delivery delay of packets} $D^p$ is the difference between the time slot in which an information packet is first transmitted by Alice and the time slot in which the packet is decoded, in order, by Bob. This is the relevant notion for real-time control applications. For frames of $k$ data information packets, the \emph{in-order delivery delay of frames} $D^f$ is the difference between the time slot in which the first information packet of a frame is transmitted by Alice and the time slot in which the $k$ packets are decoded, in order, by Bob. This is the relevant notion for video streaming applications. Finally, the \emph{in-order delivery delay of files} $D^c$ is the difference between the time slot in which Alice sends the first information packet and the time slot in which Bob is able to recover the entire file. This is the relevant notion for distributed storage applications.

We now discuss the notions of security used in this paper, illustrated in Fig.~\ref{fig: security and schemes a}. There are various notions of security in the literature; the strongest one dates back to Shannon \cite{shannon1949communication} and guarantees that no information about the message is leaked to Eve, irrespective of her computational power, i.e., $H(M|X)=H(M)$, where $H$ is the entropy, $M$ is the private message, and $X$ is the encrypted message sent through the communication network. Such a guarantee comes at a significant cost of throughput. \emph{Individual Security} \cite{bhattad2005weakly,kobayashi2013secure} achieves an optimum communication efficiency by loosening this security constraint. The way this works is by considering various messages $M_1,\ldots, M_m$ and encoding them into $X_1,\ldots, X_m$ in such a way that if Eve observes $w$ of them, it holds that $H(M_i|X_{E,w}) = H(M_i)$, for all $i = 1, \ldots, m$. In other words, Eve learns nothing about any individual message -- she only learns meaningless information about the combination of the messages.

Since both strong and individual security are information-theoretic notions, they do not depend on any limitation on Eve's computational power, and are thus post-quantum secure. However, they rely on the assumption that Eve does not observe all messages being sent between Alice and Bob, which we refer to as a \emph{weak Eve}. Thus, they provide no security against a \emph{strong Eve}, i.e., one which observes all messages.

In order to obtain security against a strong Eve, one must assume that her computational power is limited \cite{rivest1978method}. Such security schemes are referred to as \emph{computationally secure}, and rely on the conjecture that certain one-way functions are hard to invert \cite{kaltz2008introduction}. One way to achieve this is through \emph{public-key cryptography}. A public-key cryptosystem consists of an encryption function $\mathrm{Enc}(\cdot)$, a decryption function $\mathrm{Dec}(\cdot)$, a secret key $s$, and a public key $p$. Both keys are generated by Bob. The secret key is then kept securely by him, while the public key is shared to Alice, but assumed to be known to Eve. Alice encrypts the private message $M$ into $\mathrm{Enc}(M,p)$ before sending it. Bob uses the secret key to decrypt the encrypted message $M = \mathrm{Dec}(\mathrm{Enc}(M,p),s)$. The key property is that decrypting the encrypted message without the secret key is computationally expensive.

One of the first, and most widely used public-key cryptosystem is the Rivest–Shamir–Adleman (RSA) cryptosystem \cite{rivest1978method}. The security of this system relies on the computational hardness of what is known as the RSA problem which can be shown to be at least as easy as integer factorization. In \cite{365700}, however, a polynomial-time quantum algorithm for integer factorization, known as Shor's algorithm, was presented, showing that sufficiently large quantum computers can be used to break the RSA cryptosystem \cite{shor1999polynomial}. This has led to a subsequent increase in the interest of cryptosystems resilient to quantum attacks, a field known as \emph{post-quantum cryptography} \cite{bernstein2009introduction}. 

One such post-quantum cryptographic scheme is the \emph{McEliece cryptosystem} \cite{mceliece1978public}. Instead of relying on integer factorization, it relies on the problem of decoding general linear codes, which is known to be NP-Hard\cite{1055873}. The original scheme uses binary Goppa codes and suffers from low throughput, e.g., in the original paper $\eta = 0.5$, a problem which seems to be inherent to the \emph{McEliece cryptosystem} \cite{6089437}.

In order to improve the throughput, \cite{cohen2021network} introduced \emph{individual computational security}. This notion is to computational security as individual security is to strong security. Thus, as for individual security, by allowing Eve to learn only meaningless information about the combination of the messages, post-quantum security can be guaranteed with a high throughput. The secure scheme presented in \cite{cohen2021network} is known as a hybrid universal network-coding cryptosystem (HUNCC) and guarantees individual security in the case of a weak Eve, and individual computational security in the case of a strong Eve. As shown in Fig.~\ref{fig:computational_complexity}, HUNCC's computational complexity for networks is significantly lower than that of the McEliece cryptosystem.

%%%%%%%%%%%%%%%%%%%%%%%%%%%%%%%%%%%%%%%%%%%%
  \section{Setting}
%%%%%%%%%%%%%%%%%%%%%%%%%%%%%%%%%%%%%%%%%%%%

Our setting consists of a transmitter, Alice, which wants to transmit confidential data to a legitimate receiver, Bob, over $\ell$ noisy communication links, in the presence of an eavesdropper, Eve. The noise in the communication links are modeled as independent Binary Erasure Channels (BEC), i.e., the probability $\epsilon_i$ of an erasure event occurring in the $i$-th link is independent from the erasure events on the other links. We also assume that encryption via a public-key cryptosystem may only be performed on the information sent over $c\leq \ell$ communication links, and thus, cannot be performed on the remaining $w=\ell -c$ communication links.\footnote{We consider this a practical assumption that holds, for example, for the case where Alice cannot guarantee the encryption of the data before transmitting over $w$ links of the network, or for efficiency in terms of throughput and computational complexity; reducing the amount of encrypted data, reduces the high cost of applying a post-quantum cryptosystem.} As is usual for these type of networks we make use of a feedback channel.

In this paper, we focus on guaranteeing post-quantum security against a strong eavesdropper, i.e., an Eve which has access to the information transmitted over all the $\ell$ communication links, and to a quantum computer. In order to simplify the technical aspects and focus on our key techniques, we assume that the feedback channel is secure and noiseless. At each time step, $t$, Bob acknowledges Alice by sending an acknowledgment (ACK) or a negative-acknowledgment (NACK) message according to the noise realization of each link. The delay between the data transmitted at time step $t$ and the corresponding feedback is called the round trip time (RTT). We denote by $t_d$ the maximum transmission delay of a packet in bits/seconds over all the $\ell$ links, and by $t_{prop}$ the maximum propagation delay. We assume the size of the feedback acknowledgment is negligible, so that the RTT is equal to $t_d+2t_{prop}$.

% Our goal is to transmit secure, confidential data information over noisy communication links in the presence of a strong eavesdropper while minimizing in-order delivery delay and maximizing the throughput. 

%%%%%%%%%%%%%%%%%%%%%%%%%%%%%%%%%%%%%%%%%%%%
  \section{Low-Latency HUNCC} \label{sec:algos}
%%%%%%%%%%%%%%%%%%%%%%%%%%%%%%%%%%%%%%%%%%%%

\begin{figure*}[t]
    \centering
    \includegraphics[trim=0cm 0.0cm 0cm 0.0cm, width = 1 \textwidth]{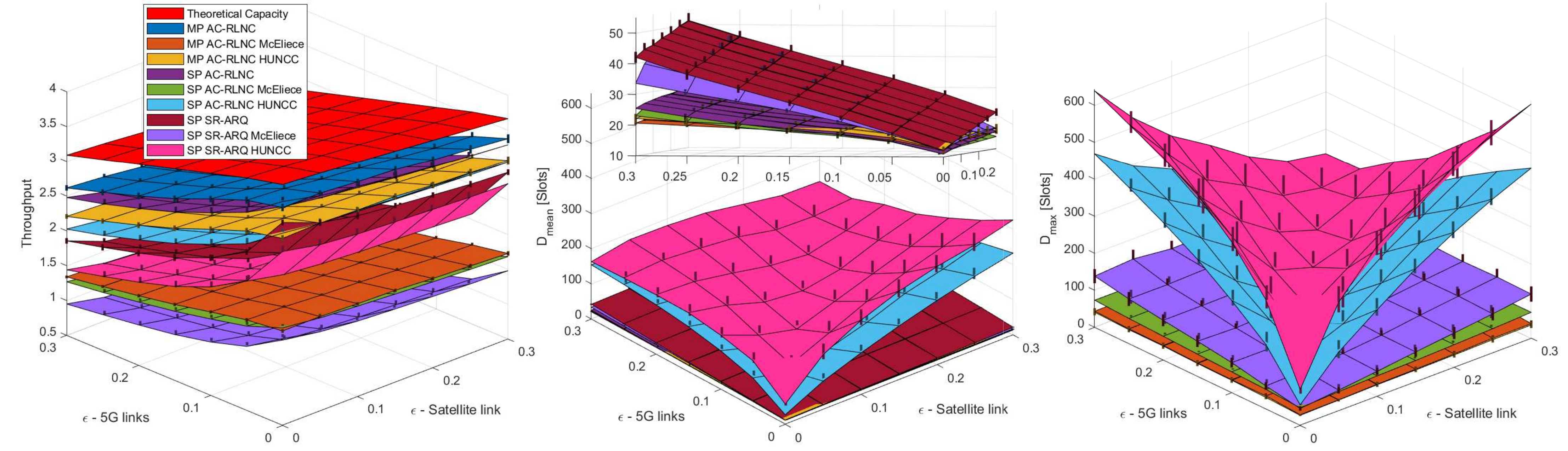}
    \caption{On the left we show the throughput for all scheme combinations. In the center and on the right, we show the mean and maximum in-order delivery delay of packets, respectively. Using HUNCC for the security stage we obtain a significantly larger throughput for all reliability schemes. However, in the case where HUNCC is combined with either SP SR-ARQ or SP AC-RLNC, this may come with a degradation of the in-order delivery delay of packets, depending on the asymmetry in the erasure realizations between the satellite link and the 5G communication links. This degradation is negligible when HUNCC is combined with MP AC-RLNC.}
    \label{fig:in_order_messages}
\end{figure*}

We propose LL-HUNCC, an efficient coding scheme with two independent stages: A pre-processing stage in which we code for security, and a second stage in which we code for reliability. Our scheme is illustrated in Fig. \ref{fig:algo_model}. In the security stage we implement HUNCC \cite{cohen2021network}, which we describe below.

Alice's messages to Bob can be represented by a string of binary bits which are partitioned into blocks, each mapped to symbols of a finite field, $M_1, \ldots, M_\ell \in \mathbb{F}_q$.\footnote{Here we assume that each $M_i \in \mathbb{F}_q$ is uniformly distributed. If that is not the case, it can be made so by utilizing the techniques in \cite{matsumoto2017universal}.} Alice then multiplies $M = (M_1, \ldots, M_\ell)$ by the generator matrix $G \in \mathbb{F}^{\ell \times \ell}_q$ of an individually secure linear code. The construction of $G$ is detailed in \cite[Section V]{cohen2021network}. We denote the output of the multiplication by $X = (X_1, \ldots, X_\ell)$. The key property guaranteed by the individually secure code $G$ is that if Eve does not observe all $\ell$ individually secure encodings $X_1, \ldots, X_\ell$, i.e., a weak Eve, then she cannot learn anything about any individual $X_i$; individual security is guaranteed. 

In order to obtain security against a strong Eve, the next step is to encrypt $c \leq \ell$ of the secure encodings $X_1, \ldots, X_\ell$ with a McEliece cryptosystem. Thus, Alice obtains the secure encoding $\Tilde{X} = (\mathrm{Enc}(X_1,p), \ldots, \mathrm{Enc}(X_c,p), X_{c+1}, \ldots, X_\ell)$. In order for a strong Eve, which is able to observe all the entries of $\Tilde{X}$, to obtain the messages $M_1, \ldots, M_\ell$, she must also break the post-quantum McEliece cryptosystem; individual computational security is guaranteed. This concludes the security stage of LL-HUNCC.

In the second stage of LL-HUNCC, and in order to obtain reliability, we choose from one of three possible approaches: SP SR-ARQ \cite{weldon1982improved}, SP AC-RLNC \cite{9076631} and MP AC-RLNC \cite{9245536}, each providing different trade-offs between security, throughput, and delay. We note that the multi-path approach does not technically satisfy the condition that encrypted packets are only sent through the trusted network. We include it, however, since we evaluate it in our simulated setting in Section \ref{sec: Evaluation and Discussion}.

Finally, after receiving enough packets, Bob is able to retrieve $\Tilde{X} = (\mathrm{Enc}(X_1,p), \ldots, \mathrm{Enc}(X_c,p), X_{c+1}, \ldots, X_\ell)$. He then uses his private key $s$ to decrypt the encoded messages $X_i = \mathrm{Dec}(\mathrm{Enc}(X_c,p),s)$, for each $i \leq c$, and obtain $X$. Via another matrix multiplication, detailed in \cite[Section V]{cohen2021network}, Bob is then able to decode the original messages $M_1, \ldots, M_\ell$.

%%%%%%%%%%%%%%%%%%%%%%%%%%%%%%%%%%%%%%%%%%%%
  \section{Performance Evaluation with Trusted Satellite and Untrusted 5G} \label{sec: Evaluation and Discussion}
%%%%%%%%%%%%%%%%%%%%%%%%%%%%%%%%%%%%%%%%%%%%

In this section we present the results of our experimental evaluation for a scenario in which we simulate a single trusted satellite communication link and three untrusted 5G communication links. We consider three performance metrics: throughput, delay, and security. We analyze the performance of the SP SR-ARQ \cite{weldon1982improved}, SP AC-RLNC \cite{9076631} and MP AC-RLNC \cite{9245536} protocols with the following security schemes: 1) No security 2) all links secured via a McEliece cryptosystem \cite{mceliece1978public} 3) HUNCC \cite{cohen2021network} with $c=1$ links, the satellite link,  encrypted by a McEliece cryptosystem. Here, the linear mixing in HUNCC is between the data transmitted across all four communication links in the network.

For each combination, we simulate three representative scenarios for typical applications to illustrate the performances of the schemes. The first scenario is for applications that requires in-order deliver of packets, e.g., real-time control solutions. The second scenario is for applications that need in-order delivery of frames (i.e., small blocks of information data), e.g., video streaming applications. The third scenarios is for in-order deliver of files, e.g., distributed storage applications.

The network setting we consider has a $RTT=20$ [slots] and erasure probabilities in each communication link varying between $0.01$ and $0.3$. The erasure probability of the satellite link varies independently from the 5G communication links, all three of which have the same erasure probability. In all simulations we consider each packet to contain 1024 bits. Hence, in the non-secure schemes, Alice transmits 1024 bits of data per packet. In the case of the McEliece cryptosystem we use a $[1024, \; 524]$-Goppa code with a normalized throughput of $\approx 0.51$, which means every $1024$ bit encrypted packet has $524$ bits of information.\footnote{This way, each encrypted packet can be unencrypted as soon as it arrives.} In the case of the HUNCC scheme, since only the satellite link is encrypted, the normalized throughput  over all four links in the network is $\approx 0.87$.

We begin by looking at the in-order delivery of packets. In Fig~\ref{fig:in_order_messages}, we show the performance of all schemes. The z-axis corresponds to throughput, the mean in-order delivery delay $D_{mean}$, and the maximum in-order delivery delay $D_{max}$, in the left, center, and right images, respectively. In all three images, the x-axis corresponds to the error rate in the satellite link and the y-axis to the error rate in the three 5G communication links. The gains in throughput and delay observed from going from the SP schemes to the MP AC-RLNC in the non-secure schemes follow from the results in \cite{9245536}. However, using the McEliece cryptosystem in every link degrades the throughput significantly. This degradation is expected from the lower normalized throughput, as discussed in the last paragraph. In contrast, this degradation is much smaller for HUNCC, since the cryptosystem is only applied over the satellite link. 

\begin{figure*}[!t]
\begin{subfigure}[]{0.52\textwidth}
    \centering
        \resizebox{\textwidth}{!}{\includegraphics[trim=5cm 0.0cm 5cm 0.0cm, clip, width = 1.2 \textwidth]{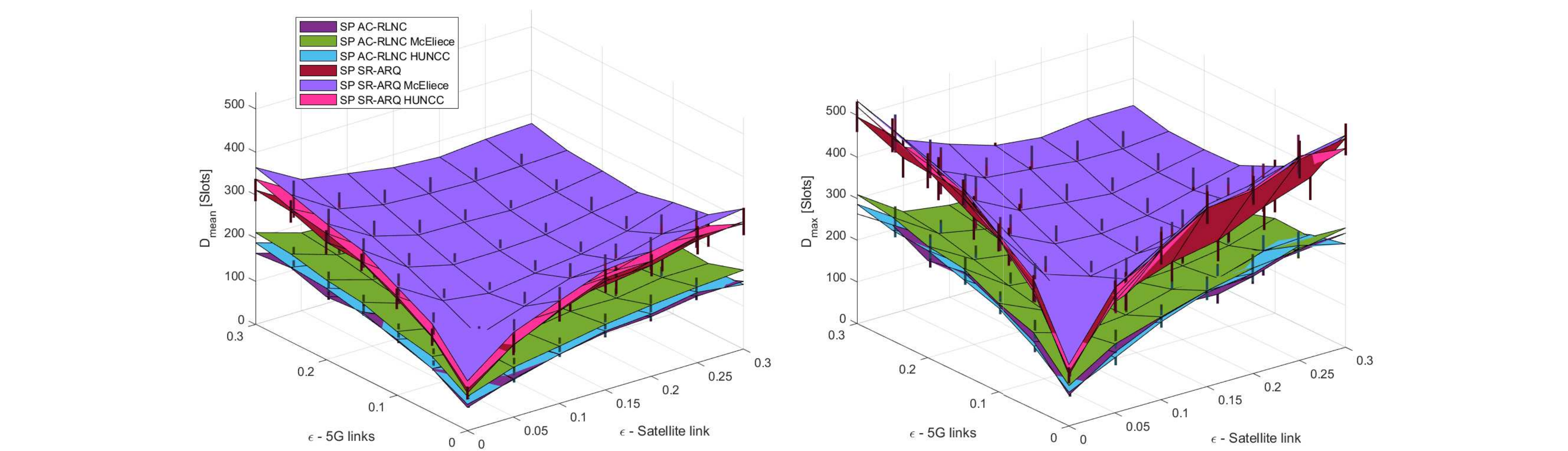}}
        \caption{In-order deliver delay of frames.}
    \label{fig:in_order_frames}
\end{subfigure}
\begin{subfigure}[]{0.52\textwidth}
    \centering
    \resizebox{\textwidth}{!}{\includegraphics[trim=5cm 0.0cm 0cm 0.0cm, clip, width = 1 \textwidth]{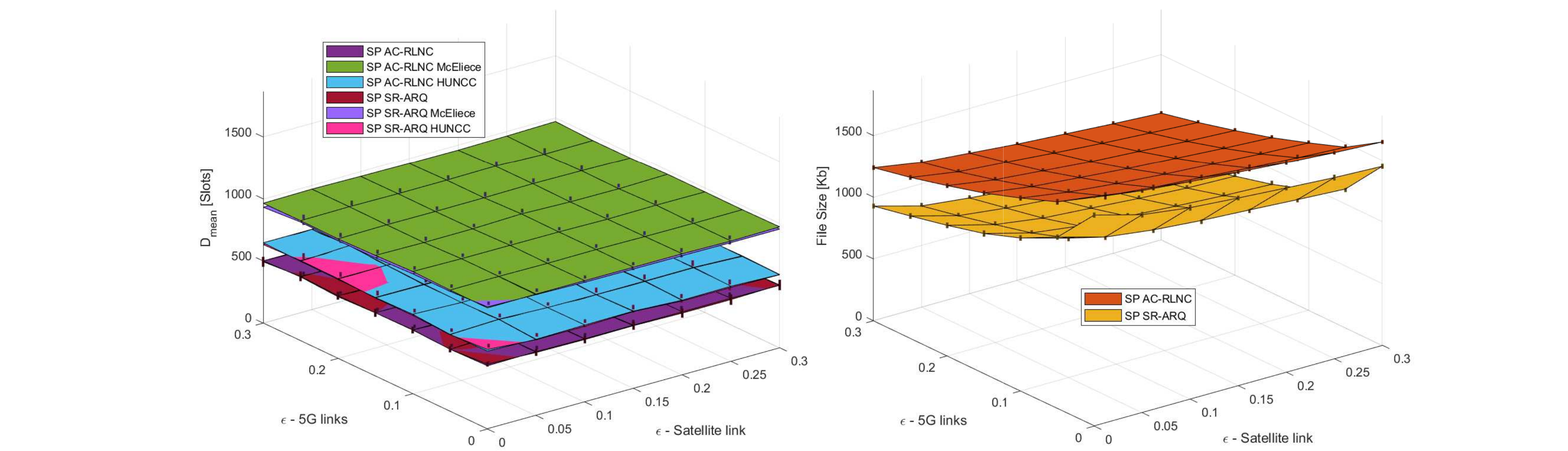}}
    \caption{In-order deliver delay of files.}
    \label{fig:completion_time}
\end{subfigure}
\caption[lof]{In \textbf{(a)} and \textbf{(b)} we show the in-order delivery delay for frames and files, respectively. The throughput for these cases remains as that of the left image on Fig.~\ref{fig:in_order_messages}. For both frames and files, the in-order delivery delay degradation is negligible when utilizing HUNCC with any of the reliability schemes.}
\label{fig: in_order_frames plus completion_time}
\end{figure*}

As for the mean and maximum in-order delivery delay (the center and right images of Fig. \ref{fig:in_order_messages}) we see the effect of the linear mixing in HUNCC when using SP AC-RLNC. In this case, in order for Bob to decode, he must first receive all relevant packets involved in the original linear combination at Alice. We note that the in-order delay of packets when HUNCC is combined with SP AC-RLNC degrades according to the asymmetry in the erasure realizations between the satellite link and the 5G communication links. When HUNCC is combined with MP AC-RLNC, however, the degradation in the in-order delivery delay is negligible, with results almost equal to that of the non-secure MP AC-RLNC scheme. That is, the mean in-order delivery delay is smaller than $20$ [Slots] and the maximum in-order delivery delay is smaller than $60$ [Slots]. This is due to the MP scheme reducing the effect of the asymmetry between the communication links. We also note that, although there is a significant degradation in throughput when the McEliece cryptosystem is used over all four links, the in-order delay per packet is almost the same as in the non-secure scheme. This is due to the independence of the packets in this scheme, i.e. Bob decrypts each packet as it is received.\footnote{Here, we do not account for the queuing time at Alice or the computation complexity of the encryption.}

We now consider the in-order delivery of frames. In Fig~\ref{fig:in_order_frames}, we show the performance of all schemes under this setting. The z-axis in the left and right images correspond to the mean and maximum in-order delivery delay of frames, $D_{mean}$ and $D_{max}$, respectively. In this scenario, the frames have a size of $100$KB. As before, each packet has $1024$ bits of information. Thus, each in-order delivery frame at the receiver contains $100$ corresponding packets in the non-secure scheme, $124$ in the HUNCC scheme, and $196$ in the pure McEliece scheme. In this case, as opposed to the in-order delivery of packets, the linear mixing in HUNCC has a negligible effect on either SP SR-ARQ or SP AC-RLNC. This occurs because the in-order delivery of frames is affected only by the last packets being sent, i.e., the delay caused by the final packets is amortized by the transmission of the initial packets.

We now look at the in-order delivery of files. In Fig~\ref{fig:completion_time} we see the average performance of 1000 packet transmissions over our simulated network. On the right of Fig~\ref{fig:completion_time}, we show the size of the achievable download file for SP SR-ARQ and SP AC-RLNC as a function of the network conditions. We see that the schemes with the McEliece cryptosystem need $2$ times more time slots than the non-secure schemes to download files of the same size. Around $1000$ [Slots] in the McEliece cryptosystem versus $500$ [Slots] in the non-secure schemes. The schemes with HUNCC needs around $630$ [Slots], only $1.3$ times more than the non-secure schemes.

\begin{figure}[ht]
    \centering
    \includegraphics[width = 0.895\columnwidth]{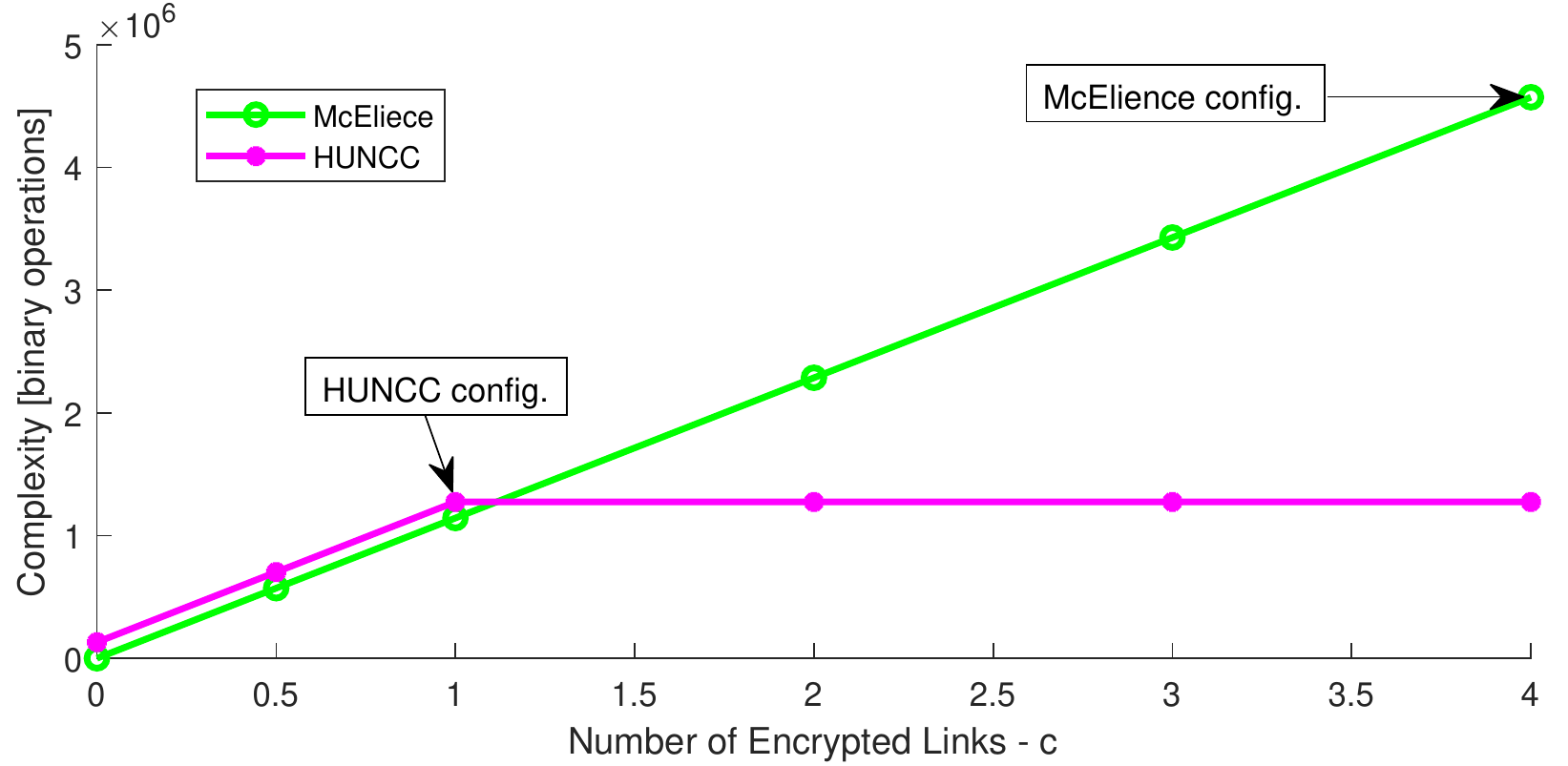}
    \caption{The computational complexity of the McEliece cryptosystem grows linearly with the amount of encrypted links, whereas HUNCC has a small overhead from the individually secure mixing, but only encrypts a single communication link.}
    \label{fig:computational_complexity}
\end{figure}

Finally, we consider the computational complexity of both HUNCC and the McEliece cryptosystem, which we illustrate in Fig.~\ref{fig:computational_complexity}. We measure this complexity by the amount of binary operations. Thus, every field operation in $\mathbb{F}_{2^m}$ costs $m$ binary operations. For the McEliece cryptosystem, we denote the length of the Goppa code by $n$ and the number of errors inserted by $t$ \cite{mceliece1978public}. Then, the normalized throughput is given by $\eta_c = 1 - \frac{tm}{2^m}$, the encryption procedure costs $\frac{\eta_ctmn}{2}$ binary operations, and the decryption costs $(3-2 \eta_c)tmn$ binary operations. For HUNCC, both the encoding and decoding consist in multiplying two $\ell \times \ell$ matrices \cite{cohen2021network}. In our simulations we considered the original McEliece parameters of $n=1024$, $m=10$, and $t=50$, obtaining $\eta_c \approx 0.51$. The encryption cost is then $131000$ binary operations and the decryption costs $1012000$ binary operations. As for HUNCC, the encoding and decoding complexity using the standard matrix multiplication algorithm requires $131072$ binary operations.

%%%%%%%%%%%%%%%%%%%%%%%%%%%%%%%%%%%%%%%%%%%%
\section*{Acknowledgements} 
%%%%%%%%%%%%%%%%%%%%%%%%%%%%%%%%%%%%%%%%%%%%
This material is based upon work supported by the USD NON-LINE under Air Force Contract No. FA8702-15-D-0001. Any opinions, findings, conclusions or recommendations expressed in this material are those of the authors and do not necessarily reflect the views of the USD NON-LINE.

%%%%%%%%%%%%%%%%%%%%%%%%%%%%%%%%%%%%%%%%%%%%
\bibliographystyle{IEEEtran}
\bibliography{bibtex/references,bibtex/Ref1,bibtex/Ref2}
%%%%%%%%%%%%%%%%%%%%%%%%%%%%%%%%%%%%%%%%%%%%

%\clearpage

%\clearpage

%The complexity of McEliece \cite{sendrier2002security} in all the links vs. HUNCC with different configurations.
%    \begin{itemize}
%       \item McEliece rate: $R_c=1-tm/2^m$ . 
%        \item Encryption cost: $C_{Enc} = (R_c/2)tmn$.
%        \item Decryption cost: $C_{Dec} = (3-2R_c)tmn$.
%        \item Original McEliece parameters \cite{baldi2009ldpc}: Goppa code with length $n=1024$ and dimension $k=524$ with $m=10$, able to correct up to $50$ errors. The key size is $k\times n = 67072$ bytes, and data information rate is $\approx 0.56$. The encryption cost is $C_{Enc} = 131000$ and the decryption cost is $C_{Dec} = 1012000$. All costs are in binary operations (one field operation in $F_{2^m}$ is counted for $m$ binary operations).
%        \item Linear mixing cost of HUNCC: $C_{HUNCC} < 2nl^3=131072$
%    \end{itemize}

\end{document}